\documentclass[runningheads,a4paper]{llncs}
\usepackage{graphicx}
\usepackage{amsfonts}
\usepackage{url}
\usepackage{multirow}
\usepackage{color}
\usepackage[letterpaper=true,colorlinks,bookmarks=false]{hyperref}
\usepackage{bm}
\usepackage{authblk}
\usepackage{amssymb}
\usepackage{graphicx}
\usepackage{multirow}
\usepackage{algorithm}
\usepackage{algpseudocode}
\usepackage{makeidx}
\usepackage{multirow}
\usepackage{graphicx}
\graphicspath{{pics/}{figs/}}
\usepackage{graphics,epsfig,subfigure}
\usepackage{epstopdf}
\usepackage{amsmath} 
\usepackage{subfigure}

\title{The Channel Attention based Context Encoder Network for Inner Limiting Membrane Detection}
				\author{Hao Qiu\inst{1, 2}, 
Zaiwang Gu\inst{3},
Lei Mou\inst{2},
Xiaoqian Mao\inst{2},
Liyang Fang\inst{2},
Yitian Zhao\inst{2},
Jiang Liu\inst{3},
Jun Cheng\inst{2}\thanks{Corresponding author: Chengjun@nimte.ac.cn}
 		\  
	} 
	\institute{
		{$^{1}$School of Mechatronic Engineering and Automation, Shanghai University\\ $^{2}$ Cixi Institute of Biomedical Engineering, Ningbo Institute of Industrial Technology, Chinese Academy of Sciences
\\$^{3}$ Department of Computer Science and Engineering, Southern University of Science and Technology}
			}

\begin{document}
 
\maketitle
\begin{abstract} The optic disc segmentation is an important step for retinal image based disease diagnosis such as glaucoma. The inner limiting membrane (ILM) is the first boundary in the OCT, which can help to extract the retinal pigment epithelium (RPE) through gradient edge information to locate the boundary of the optic disc. Thus, the ILM layer segmentation is of great importance for optic disc localization. In this paper, we build a new optic disc centered dataset from 20 volunteers and manually annotated the ILM boundary in each OCT scan as ground-truth. We also propose a channel attention based context encoder network modified from the CE-Net \cite{Gu2019CENetCE} to segment the optic disc. It mainly contains three phases: the encoder module, the channel attention based context encoder module, and the decoder module. Finally, we demonstrate that our proposed method achieves state-of-the-art disc segmentation performance on our dataset mentioned above.

\end{abstract}

\section{Introduction}

 \begin{figure*}[t]
\centering{
\includegraphics[width=12cm]{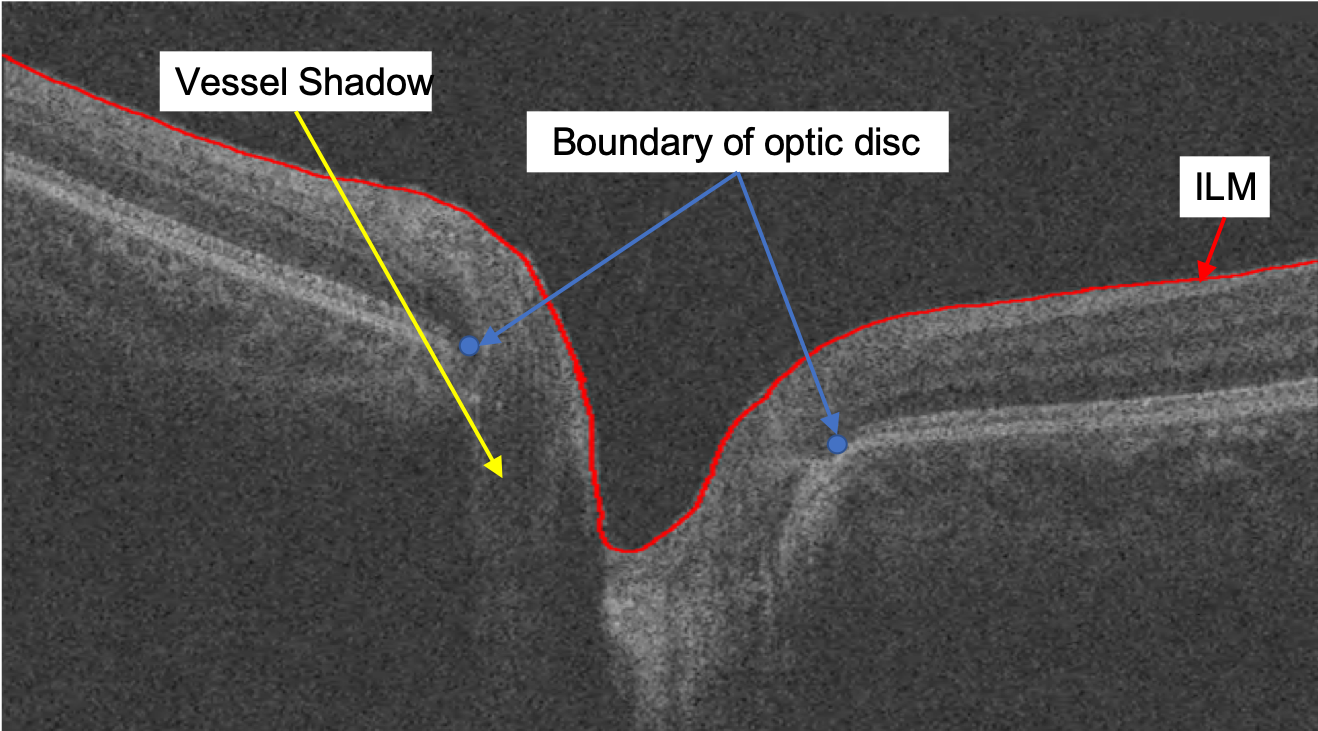}
}
\caption{Optic nerve head structure in a cropped OCT slice. The red curve denotes the ILM boundary. The blue points refer to the boundary points of the optic disc. ILM: Inner limiting membrane. }
\label{fig1}
\end{figure*}

Glaucoma is the second leading cause of blindness globally, which may result in vision loss and irreversible blindness. The number of people suffering from glaucoma is estimated to increase to 80 million in 2020 \cite{tham2014global}. As the disease progresses asymptomatic in the early stages, the majority of the patients are unaware until an irreversible visual loss occurs. Thus, early diagnosis and treatment for glaucoma is utmost essential for preventing the deterioration of vision. While there are various approaches to diagnose glaucoma such as vessel distribution, FFT/B-spline coefficients, most of the known literature has endeavoured to assess the cup-to-disc ratio (CDR) .

There have been a number of attempts at automatically detecting the optic disc in ocular images. Many proposed optic disc detection approaches concentrate on segmenting the optic region in color fundus images. For example, Liu \emph{et al.}  \cite{liu2013automatic} proposed Variational level set approach for segmentation of optic disc without reinitialization. Xu \emph{et al.} \cite{xu2007optic} employed the deformable model technique through minimization of an energy function to detect the disc. Cheng \emph{et al.} \cite{cheng2015sparse} used the state-of-the-art self-assessed disc segmentation method combined three methods to segment the disc. However, these proposed approaches face challenges when the optic disc does not have a distinct color in the fundus image.

\begin{figure*}[t]
\centering{
\includegraphics[width=12.5cm]{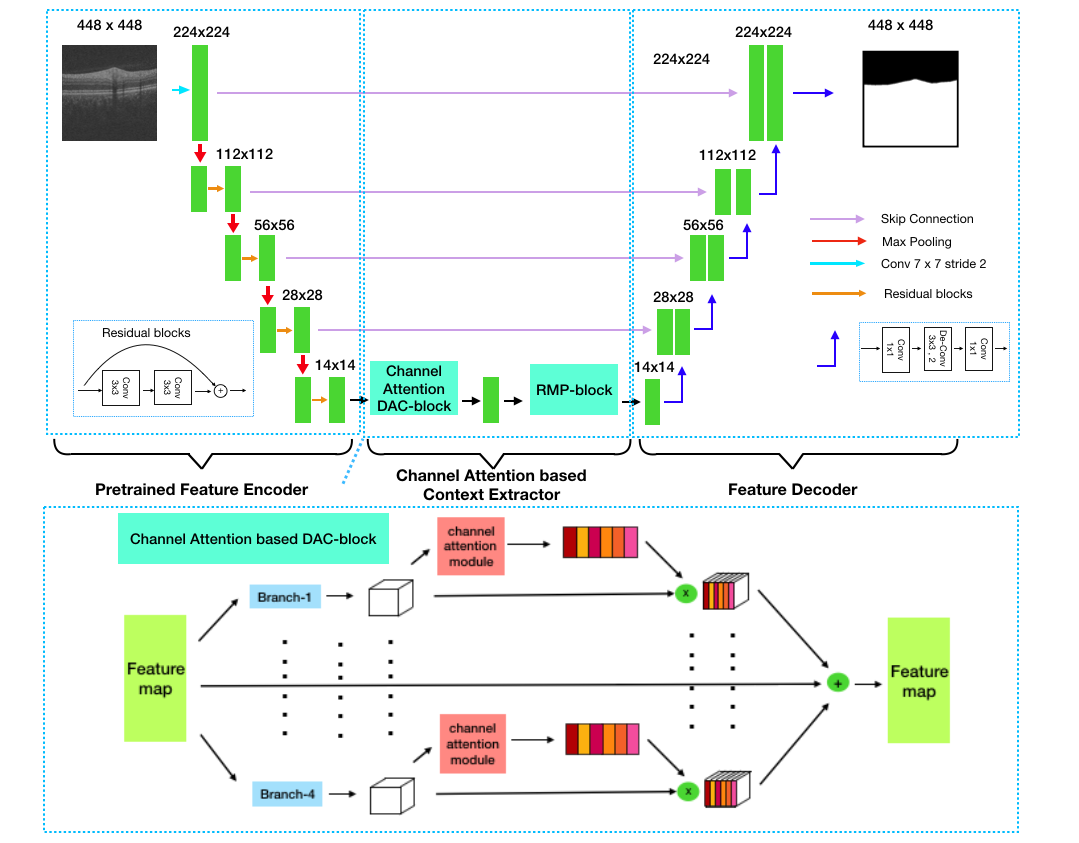}
}
\caption{Illustration of the proposed CACE-Net. Firstly, the images are fed into a feature encoder module, where the residual network (ResNet) block was employed as the backbone for each block, and then followed by a max-pooling layer to increase the receptive field for better extraction of global features. Then the features from the encoder module are fed into the proposed channel attention based context encoder module. Finally, the decoder module was used to enlarge the feature size and output a mask, the same size as the original input. }
\label{fig2}
\end{figure*}

Optical coherence tomography (OCT) , an important retinal imaging method with non-invasive, high-resolution characteristics, provides the fine structure within the human retina \cite{schmitt1999optical}. A single image of OCT slice is shown in Fig. \ref{fig1}. Some optic disc segmentation methods are applied to 3-D OCT volumes. For example, Lee \emph{et al.} \cite{lee2017deep} applied a K-NN classifier to segment the optic disc cup and neuroretinal . Fu \emph{et al.} \cite{fu2014automatic} provided a Low-rank reconstruction to automatically detect optic disc in OCT slices.

With the development of convolutional neural network (CNN) in image and video processing \cite{krizhevsky2012imagenet}, automatic feature learning algorithms using deep learning have emerged as feasible approaches and are applied to handle the  image analysis. Recently, some deep learning based segmentation algorithms have been proposed to segment medical images \cite{gu2018deepdisc}, \cite{Gu2019CENetCE}. Based on the U-Net, a recent popular medical image segmentation architecture, CE-Net employs multi-scale atrous convolution and pooling operations to improve the segmentation performance. And it achieves some state-of-the-art performance in some medical image segmentation tasks, such as optic disc segmentation and OCT layers segmentation. 
The original context extractor module in CE-Net was consist of a dense atrous convolution (DAC) module and a residual multi-kernel pooling (RMP) module. However, the original DAC and RMP accounted for abundant channels to enrich the semantic features representations. Each channel of the features at the classification layer can be regarded as a speciﬁc-class response since we add the supervision signal on this layer. These abundant channels could be further embedded to produce the global distribution of channel-wise feature responses. In this paper, in order to extract more high-level semantic features, we introduce the channel attention mechanism to enhance the context extractor module of the CE-Net, and propose a channel attention based context encoder network (called CACE-Net) for inner limiting membrane detection.

The major contributions of this work are summarized as follows: 

1) We annotate 20 3D-OCT scans (both of them are right eye scans) centered at optic disc.

2) we leverage the ability of CACE-Net to accurately segment the inner limiting membrane (ILM) in our dataset, which is defined as the boundary between the retina and the vitreous body. This is necessary for our further work to detect the optic disc boundary points. The segmentations on database of OCT images are demonstrated to be superior to those from some known state-of-the-art methods. And we will release our code and dataset on Github later.

\section{Proposed Method}

The CE-Net \cite{Gu2019CENetCE} achieves the state-of-the-art performances in some 2D medical image segmentation tasks, such as optic disc segmentation, retinal vessel detection, lung segmentation and cell contour extraction. The proposed CACE-Net is modified from the CE-Net, which mainly contains three phases: the encoder module, the channel attention based context encoder module, and the decoder module, as shown in Fig. \ref{fig2}. The feature encoder module includes four encoder blocks, and the residual network (ResNet) block was employed as the backbone for each block, and then followed by a max-pooling layer to increase the receptive field for better extraction of global features. Then the features from the encoder module are fed into the proposed channel attention based context encoder module. Finally, the decoder module was used to enlarge the feature size and output a mask, the same size as the original input. 

\subsection{Channel attention based context extractor module} 

The original context extractor module in CE-Net \cite{Gu2019CENetCE} employed four cascade branches with multi-scale atrous convolution to capture multi-scale semantic features, followed by various size pooling operations to further encode the multi-scale context features. This module accounts for abundant channels to enrich the semantic features representations, which could be further embedded to generate the global distribution of channel-wise feature responses. Therefore, motivated by the SE-Net \cite{hu2018squeeze}, we propose a channel attention based context extractor module, introducing the relationship between channels.

In this section, we mainly introduce how to exploit the interdependencies of channel maps, as illustrated in Fig. \ref{fig2}. The proposed channel attention based context extractor module employs channel attention mechanism to allow the network to perform feature recalibration of aggregated context features, with the basis of original DAC block. Specially, the CACE module utilizes four cascade branches with multi-scale atrous convolution and channel attention module, to gain high-level features. 

\begin{figure*}[t]
\centering{
\includegraphics[width=12cm]{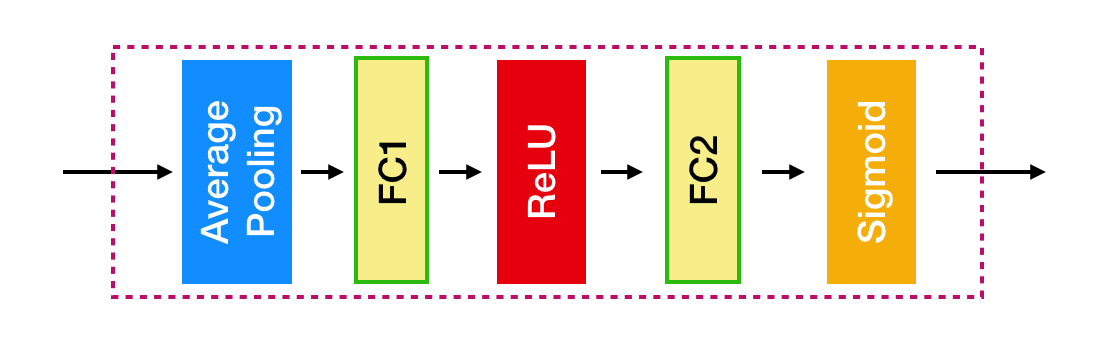}
}
\caption{Illustration of the channel attention module. }
\label{ca}
\end{figure*}

As illustrated in Fig. \ref{ca}, the extracted feature map $F \in \mathbb{R}^{C\times H\times W}$ in channel attention module is first calculated directly by the global average pooling to generate channel-wise statistics $z \in \mathbb{R}^{C}$ :

\begin{equation}
\label{eq1}
z_{c} = \frac{1}{H\times W}\Sigma_{i=1}^{H}\Sigma_{j=1}^{W}f_{c}(i,j)
\end{equation}
where $H \times W$ represents the spatial dimensions of features and $C$ is the number of channels. Then, the two linear transformations $W_{1}, W_{2}$ and a sigmoid activation function $\sigma$ are employed to obtain the squeeze and excitation statistics $s \in \mathbb{R}^{C}$:
\begin{equation}
\label{eq1}
s_{c} = \sigma(W_{2}\delta(W_{1}z_{c}))
\end{equation}
where $\delta$ refers to the ReLU function, $W_{1} \in \mathbb{R}^{\frac{C}{r}\times C}$ and $W_{2} \in \mathbb{R}^{C \times \frac{C}{r}}$. Finally, a matrix multiplication between the statistics $s \in \mathbb{R}^{C}$ and the feature $F \in \mathbb{R}^{C\times H\times W}$ is added to obtain the final output in each branch of the proposed channel attention DAC module, followed by the RMP block for further context information with multi-scale pooling operations.

\subsection{Feature Decoder}

Instead of directly upsampling the features to the original image dimensions, we follow the CE-Net \cite{Gu2019CENetCE} to introduce a feature decoder module that restores the dimensions of the high level semantic features layer by layer. In each layer, we use ResNet block as the backbone of the decoder block which is followed by a 1 $\times$ 1 convolution, a 3 $\times$ 3 transposed convolution, a 1 $\times$ 1 convolution. Similar to U-Net \cite{ronneberger2015u} , we add a skip connection between each layer of the encoder and decoder. Finally, the feature decoder module could generate the prediction of the same size as the original input. 

\subsection{Boundary Extractor}

The main goal of this method is to detect internal limiting membrane. Therefore, we need to turn the segmentation prediction to a boundary line, which corresponds to the internal limiting membrane. We remove the small connected components to denoise the segmentation prediction, adopting the morphology method. After this post processing operation, we achieve the final boundary corresponding to the internal limiting membrane between the retina and the vitreous body. 

\subsection{Loss Function}
In this method, we choose binary cross-entropy loss as our loss function $\mathcal{L}_{B}$, since the method just needs to predict the binary outputs. The binary cross-entropy loss is as follows: 
\begin{equation}
	\mathcal{L}_{B}= -\mathbb{E}_{\bm{x}\sim p_{data}}[\bm y\cdot \log(D(\bm x))+(1- \bm{y})\cdot \log(1-D(\bm x))],
\end{equation}
where $\bm y$ represents the ground truth, and $D(\bm x)$ is the prediction.

\section{Experiment Results}

\subsection{Dataset and metric}

20 3D-OCT scans (both of them are right eye scans) centered at optic disc were collected from 20 volunteers. Each OCT scan consisted of $885\times512$ image resolution. while there exist methods for extracting multiple retinal layers from OCT slices, only ILM layer boundaries is needed in our paper. The ILM is defined as the boundary between the retina and the vitreous body, which is the first boundary of retinal OCT. The ground-truth optic disc boundary of a 3D-OCT volume is obtained by first manually labeling the optic disc points in each optic disc centered slice (with a trained labeler and two experts for quality control) . These labeled points were then to generate the ground-truth optic disc boundary. In our paper, we also randomly take 10 people's images for training, and others for testing. In this paper, we follow the same partition of the data set to train and test our models.

Following the previous approaches \cite{Gu2019CENetCE}, we compute the mean absolute error (mae) between prediction and ground truth as the metric to evaluate the accuracy of segmentation algorithms. 
\begin{equation}
    \label{metric}
error = \frac{1}{n}\sum_{i=1}^{n}|y_{i}-Y_{i}|
\end{equation} 
where $y_{i}$ represents the $i_{th}$ pixel predicted value of one surface, and $Y_{i}$ represents that of ground truth.

\subsection{Implementation details}

The proposed CACE-Net was implemented on PyTorch library with the NVIDIA GPU. We choose stochastic gradient descent (SGD) optimization, other than adaptive moment estimation (Adam) optimization. We use SGD optimization since recent studies \cite{keskar2017improving} show that SGD often achieves a better performance, though the Adam optimization convergences faster.  The initial learning rate is set to 0.001 and a weight decay of 0.0001. We use poly learning rate policy where the learning rate is multiplied by $\left( 1- \frac{iter}{max\_iter}\right)^{power}$ with power 0.9. All training images are rescaled to $448 \times 448$.

\begin{figure*}[t]
\centering{
\includegraphics[width=12cm]{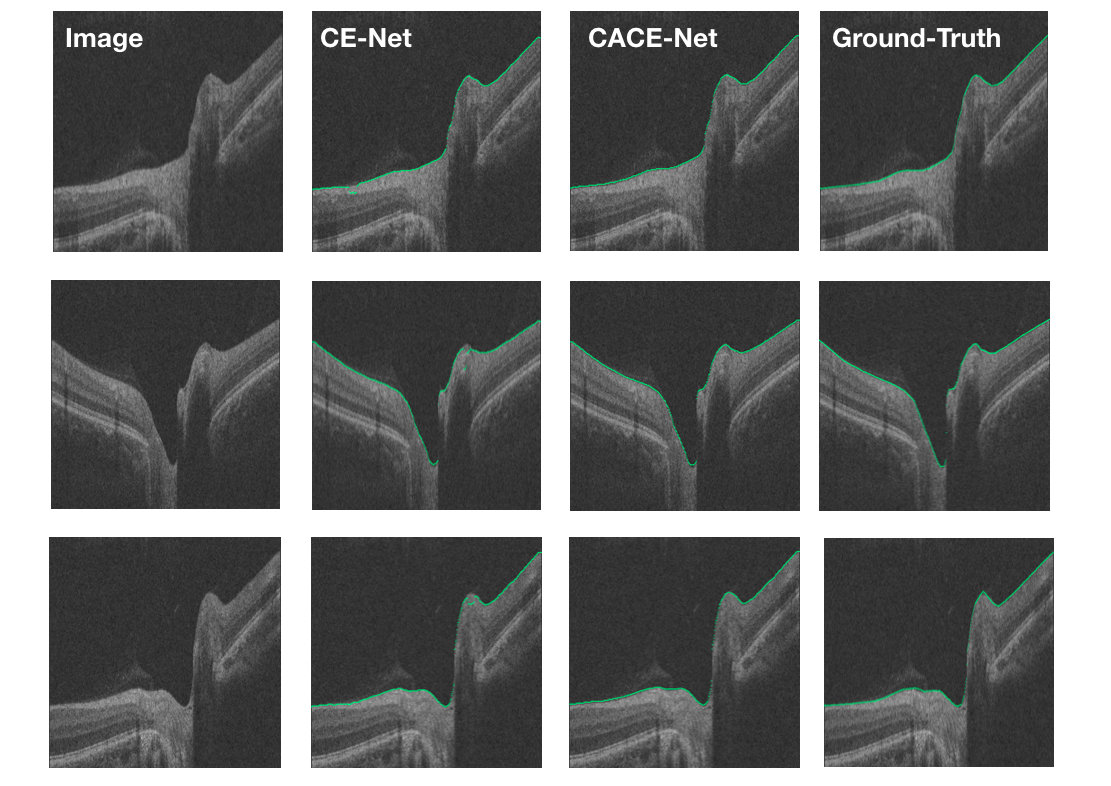}
}
\caption{Sample results of the ILM segmentation. From left to right: original images, CE-Net, CACE-Net and Ground-Truth }
\label{result} 
\end{figure*}

In order to demonstrate conclusively the superiority of the proposed method over the other methods, we compare our method with two algorithms for the ILM segmentation:

(1) U-net, a popular neural network architecture for biomedical image segmentation tasks.

(2) CE-Net \cite{Gu2019CENetCE}, which achieves the state-of-the-art performances in some 2D medical image segmentation tasks, such as optic disc segmentation, retinal vessel detection,lung segmentation and cell contour extraction.

\subsection{Results and discussion}
\begin{table}[]
\caption{PERFORMANCE COMPARISON OF THE ILM DETECTION (MEAN $\pm$ STANDARD DEVIATION)}
\label{tab:my-table}
\centering
\begin{tabular}{ccccc}

\cline{1-4}
\multicolumn{1}{c|}{Method} & \multicolumn{1}{c|}{U-Net} & \multicolumn{1}{c|}{CE-Net}        &  CACE-Net          &  \\ \cline{1-4}
\multicolumn{1}{c|}{$error$}      & \multicolumn{1}{c|}{6.404$\pm$16.407} & \multicolumn{1}{c|}{2.467$\pm$1.989} & 2.199$\pm$1.471 &  \\ \cline{1-4}
                           &                             &                         &            &  \\
                     
\end{tabular}
\end{table}

As can be seen in Table \ref{tab:my-table}, we show the performances of three optic disc segmentation algorithms.  Compared with other state-of-the-art optic disc segmentation methods, our CACE-Net  outperforms the other algorithms based on deep learning image processing method. From the comparison shown in Table \ref{tab:my-table}, the CACE-Net achieves 2.199 in the mean absolute error, better than the U-Net. From the comparison between CE-Net \cite{Gu2019CENetCE} and our CACE-Net, we also observe that there is a drop of the mean absolute error by 10.8\% from 2.467 to 2.199.

We also show three sample results in Fig. \ref{result} to visually compare our method with the most competitive methods, CE-Net. The comparison images show that our method obtain more accurate segmentation results.

\section{Conclusion}

In this paper, we have built a manually labeled OCT dataset and proposed an effective architecture for segmenting the ILM layer in our OCT dataset. The proposed CACE-Net achieves the mean absolute error of 2.199 in our dataset, better than other methods.

\bibliographystyle{splncs}
\bibliography{paper31}

\end{document}